\documentclass[a4paper]{jpconf}
\usepackage[pdftex]{graphicx}
\usepackage{amsmath,amssymb,color,hyperref,graphicx,pdfsync,overpic,color,rotating,dashrule,float,bm,multicol,setspace,cancel}
\usepackage[table]{xcolor}
\usepackage{tabularx,booktabs}
\usepackage{mathrsfs}
\usepackage{mathtools}
\usepackage{cancel}
\usepackage{epsfig,epstopdf}
\usepackage{cite}

\newcommand{\bsa}{\boldsymbol{a}}

\newcommand{\bsf}{\boldsymbol{f}}

\newcommand{\bsx}{\boldsymbol{x}}

\newcommand{\bsU}{\boldsymbol{U}}
\newcommand{\bsA}{\boldsymbol{A}}

\newcommand{\bsY}{\boldsymbol{Y}}
\newcommand{\bsu}{\boldsymbol{u}}

\usepackage{graphicx}
\usepackage{grfext}

\ifx\mydraft Y
  \usepackage{epstopdf-base}
  \csname define@key\endcsname{Gin}{pdfpng}[]{%
    \epstopdfDeclareGraphicsRule{.pdf}{png}{.png}{%
      convert ##1 \OutputFile
    }%
  }
\else
  \PrependGraphicsExtensions*{.pdf}
  \csname define@key\endcsname{Gin}{pdfpng}[]{}
\fi

\graphicspath{{Figures/}}

\begin{document}
\title{Linear and nonlinear Granger causality analysis of turbulent duct flows}
\author{B Lopez-Doriga$^1$, 
M~Atzori$^2$,
R~Vinuesa$^3$, 
H~J~Bae$^{4}$,
A~Srivastava$^{1}$,
and S~T~M Dawson$^1$}
\address{$^1$Mechanical, Materials \& Aerospace Engineering Department, Illinois Institute of Technology, Chicago IL 60616, USA}
\address{$^2$Dipartimento di Scienze e Tecnologie Aerospaziali, Politecnico di Milano (DAER), Milano 20156, Italy}
\address{$^3$FLOW, Engineering Mechanics, KTH Royal Institute of Technology, Stockholm 10044, Sweden}
\address{$^4$Graduate Aerospace Laboratories, California Institute of Technology, Pasadena CA 91125, USA}
\ead{blopezdorigacostales@hawk.iit.edu}
\begin{abstract}
This research focuses on the identification and causality analysis of coherent structures that arise in turbulent flows in square and rectangular ducts. Coherent structures are first identified from direct numerical simulation data via proper orthogonal decomposition (POD), both by using all velocity components, and after separating the streamwise and secondary components of the flow. The causal relations between the mode coefficients are analysed using pairwise-conditional Granger causality analysis. We also formulate a nonlinear Granger causality analysis 
that can account for nonlinear interactions between modes. Focusing on streamwise-constant structures within a duct of short streamwise extent, we show that the causal relationships are highly sensitive to whether the mode coefficients or their squared values are considered, whether nonlinear effects are explicitly accounted for, and whether streamwise and secondary flow structures are separated prior to causality analyses. We leverage these sensitivities to determine that linear mechanisms underpin causal relationships between modes that share the same symmetry or anti-symmetry properties about the corner bisector, while nonlinear effects govern the causal interactions between symmetric and antisymmetric modes.  In all cases, we find that the secondary flow fluctuations (manifesting as streamwise vorticial structures) are the primary cause of both the presence and movement of near-wall streaks towards and away from the duct corners.
\end{abstract}

\section{Introduction}
There exist a variety of modal analysis methods that can be used to identify coherent structures in fluid flows \cite{taira2017modal,taira2020modal}. However, additional analysis is often required to understand the causal mechanisms that govern the formation and interaction between such structures, particularly true for data-driven decomposition methods such as the proper orthogonal decomposition (POD). 

A number of recent studies have introduced quantitative causality analyses for fluid mechanics problems \cite{tissot2014granger,lozanod2020causalityeddies,lozanod2021causalitywall,lozanod2022causality,martinez2022causality}. Among numerous frameworks, Granger causality \cite{granger1969} represents a relatively simple and robust scheme, although its original formulation is limited to linear interactions between the time series of different variables/quantities. However, this linear formulation enables direct connections with other linear methods used in fluids, such as dynamic mode decomposition \cite{gunjal2023granger}, which can be used to model the underlying linear dynamics assumed to describe the system for such analysis. A major drawback of Granger causality is its linearity, which may render it insufficient to expose causal behavior for nonlinear systems. On the other hand, transfer entropy causality analyses \cite{schreiber2000} have successfully identified relevant dynamics in the field of fluids \cite{bae2018causality,lozano2020causality,wang2021transferEnt}. One of the main strengths of transfer entropy is that it is agnostic to the form of the system dynamics, although a significantly large amount of data and a high computational cost are required to achieve converged results \cite{knearest1987,kraskov2004,duan2013}. Here, we opt for Granger causality analysis and extend its scope to include second-order quadratic interactions. This makes the framework more amenable for the study of nonlinear systems
while simultaneously exploiting the advantages and simplicity of the method. Furthermore, the proposed method will enable disambiguation between linear and nonlinear causal mechanisms.

This work focuses on turbulent flows through a square duct. This flow configuration has been the subject of a number of prior investigations \cite{pinelli2010,vinuesa2014,vinuesa2015minimum,vinuesa2018secondary,matin2018coherent,pirozzoli2018,orlandi2019,modesti2019direct,khan2020dynamics,lopez2022resolvent}. 
One characteristic feature that distinguishes such flows from infinitely-wide channel flow is the presence of secondary mean flow components perpendicular to the direction of the bulk flow \cite{nikuradse1930,prandtl1931}, which typically features a pair of counter-rotating streamwise vortices in each corner, and therefore increases the streamwise velocity near the corner. Moreover, the configuration of this secondary flow does not settle until $Re_\tau \approx 300$ \cite{gavrilakis_2018}, although all the intermediate configurations share a similar pattern that involves counter-rotating streamwise vortices. While it has been postulated for several decades that this secondary flow emerges due to components of the Reynolds stress \cite{gessner1973origin,vinuesa2014}, there remains uncertainty as to how the dynamics and causal relationships between near-wall coherent structures contribute to this phenomena. This work will seek to advance our understanding of such mechanisms.

The paper is structured as follows. Section~\ref{sec:description} provides a description of the main features of the turbulent duct flow. This is followed by an overview of POD in Section~\ref{sec:pod}. Section~\ref{sec:causality} presents the linear/standard formulation of Granger causality analysis and introduces a nonlinear formulation of the framework. The results of the present investigation are divided into: Section~\ref{sec:moehlis}, where we discuss the accuracy of linear and nonlinear Granger causality analyses on the low-order turbulence model introduced by Moehlis et al.~\cite{moehlis2004}, and compare with the results presented in Mart{\'\i}nez-S{\'a}nchez et al.~\cite{martinez2022causality}; and Section~\ref{sec:results_pod}, in which we examine the coherent structures identified by POD of the turbulent data and discusses the results obtained by both linear and nonlinear Granger causality analyses. Lastly, Section~\ref{sec:conclusions} reviews the main findings of this work and discusses future prospects. 
  
\section{Turbulent duct flow}
\label{sec:description}
We focus on  flow through a square duct, governed by the incompressible Navier--Stokes equations
\begin{equation}
\label{eq:momentum}
    \frac{\partial \bsu}{\partial t} = -\bsu \cdot \nabla\bsu-\nabla p+\frac{1}{Re}\Delta\bsu,
\end{equation}
\begin{equation}
\label{eq:cont}
    \nabla\cdot\bsu=0,
\end{equation}
where $\bsu=[u(\bsx,t),v(\bsx,t),w(\bsx,t)]^T$ represents the instantaneous velocity field, $\bsx=(x,y,z)^T$ are the spatial coordinates and $p=p(\bsx,t)$ denotes pressure. These enforce conservation of momentum and continuity, respectively. 
Here $x$ denotes the streamwise direction, and the other two dimensions ($y$, $z$) are confined to the rectangular domain $\Omega=\lbrace y \in [-h,h]\rbrace \times \lbrace z \in [-L,L]\rbrace$, where $A=L/h$ is the aspect ratio of the rectangular cross-section. This study will only consider the case $A=1$. The gradient operator is given by $\nabla = [\partial_x,\partial_y,\partial_z]^T$ and the Laplacian operator is defined as $\Delta = \nabla \cdot \nabla$. No-slip and no-penetration conditions are imposed at all wall-boundaries. In the numerical simulations, periodic boundary conditions are applied in the streamwise direction. Note that velocity is nondimensionalized by the friction velocity $u_\tau = \sqrt{\tau_w/\rho}$, with $\tau_w$ denoting the wall-shear stress, and the friction Reynolds number is defined as $Re_\tau = h u_\tau/\nu$, where $\nu$ corresponds to the kinematic viscosity.

\begin{figure}[ht!]
\vspace*{-0.2cm}
\centering {
{\hspace*{-0.cm}\includegraphics[width= 0.9\textwidth]{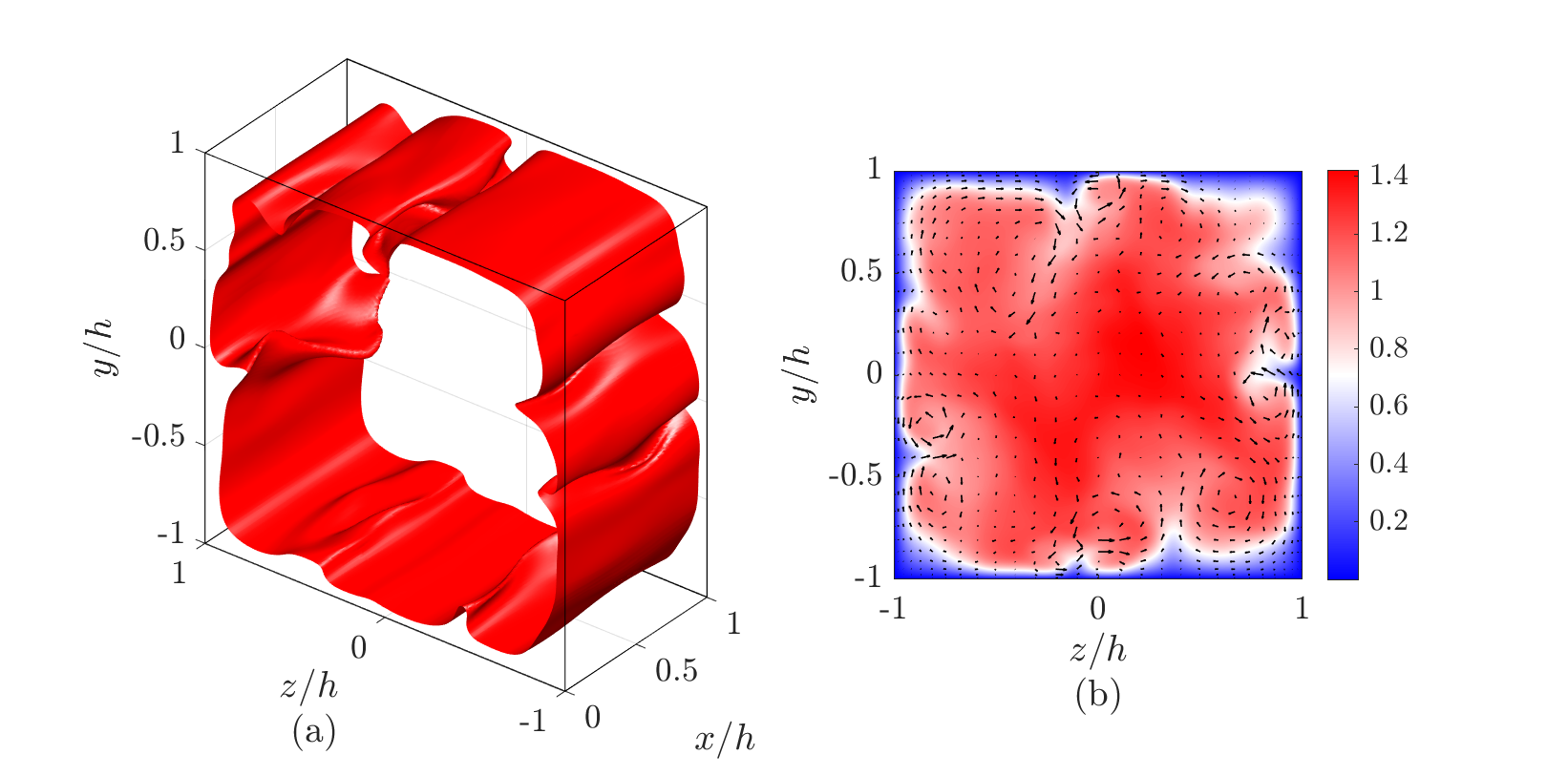}} }
\vspace*{-0.cm}
\caption{(a) Contour level of a sample snapshot of the instantaneous streamwise velocity component $u(x,y,z,t)$ and (b) streamwise component and vector field of the $(y,z)$ components of the streamwise average of a single snapshot $\bsu(\overline{x},y,z,t)$ for a minimal turbulent duct, with $Re_\tau=180$ and $A=1$.
}
\vspace*{-0.cm}
\label{fig:2snapshots}
\end{figure} 

\begin{figure}[ht!]
\vspace*{-0.1cm}
\centering {
{\hspace*{-1.4cm}\includegraphics[width= 1.2\textwidth]{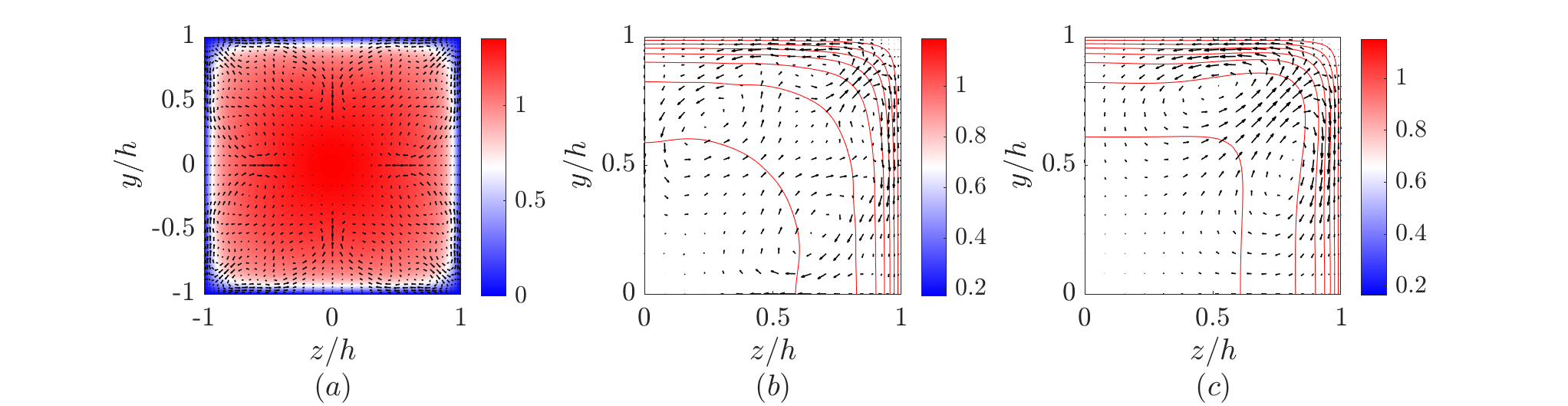}} }
\vspace*{-0.3cm}
\caption{(a) Filled contour plot of the streamwise component and vector field of the secondary components of the turbulent mean $\bsU(y,z)$ of the minimal (short) duct;  (b)  detail of the mean $\bsU(y,z)$ in the upper-right corner of the minimal duct, and (c) the equivalent plot for the long duct, all  with $Re_\tau=180$ and $A=1$.
}
\vspace*{-0.2cm}
\label{fig:turbulentMean}
\end{figure}

The majority of the analysis will consider data from a short direct numerical simulation of flow through a square duct of streamwise length $h$, and cross-section $2h \times 2h$. We use 4,000 snapshots of data, collected with at a friction Reynolds number $Re_\tau = 180$ with a time-step $\Delta t=1$ (nondimensionalized by the bulk velocity $U_b$ and the lengthscale $h$). Figure~\ref{fig:2snapshots}~(a) shows a representative snapshot of this system. Further details regarding this configuration, and the numerical methods used for the DNS, can be found in Refs.~\cite{minimalDuct2018,atzori2021intense}. Due to the very small streamwise extent of this particular configuration, we refer to this case as the short duct. 
While the short streamwise extent eliminates some of the phenomena and complexity present in a longer duct, it still retains some of its key features. In particular, the mean secondary flow includes two counter-rotating vortices in each of the corners of the duct, oriented such that mean flow is directed from the duct center towards the corner. 
As a result, the streamwise mean component features non-convex contour levels near these regions \cite{nikuradse1926}. Note that while some of these vortices can be identified on the individual snapshots (see Figure~\ref{fig:2snapshots}~(b)), the four pairs are rarely identified in one single instance, 
and in fact only appear altogether when the temporal mean is computed. This secondary mean flow, usually referred to as Prandtl's secondary flow of the second kind \cite{prandtl1931}, is shown in Figure \ref{fig:turbulentMean}. In addition to showing the mean velocity for the short duct configuration, in Figure \ref{fig:turbulentMean}(c) we also show the mean flow for a much longer duct, with streamwise length of $25h$ (the corresponding streamwise mean flow is not shown here since there are no significant differences with respect to Figure~\ref{fig:turbulentMean}(a)). While the magnitude of the secondary flow is larger for the longer duct (reaching a maximum amplitude of approximately $2\%$ of the streamwise mean, in comparison to approximately $0.4\%$ for the short duct), we observe that both configurations share secondary flow profiles with the same key features. As such, this short duct configuration represents a simplified system that can be used both to test the causality analysis methods that are developed in the present work, and to obtain insight into the physical mechanisms underlying turbulent duct flow.

\section{Proper orthogonal decomposition}
This section provides a  description of the modal decomposition technique used in this paper to extract the dominant coherent structures that arise within turbulent duct flows, which will be the subject of the causality analysis performed in the second part of this study.
\label{sec:pod}
POD \cite{lumley1967pod,berkooz1993pod,karhunen1946pod,loeve1955prob} identifies modal spatial structures from a vector field in a manner such that each of the coherent modes captures an optimal fraction of the total energy content. In fact, the term \textit{proper} or optimal alludes to the fact that the the kinetic energy captured by the first $n$ spatial modes is maximized, providing a low-order representation of the field of interest. 
In this work, this technique will naturally apply to turbulent flows, although it has been fruitful in several other applications \cite{georgiou1999,kappagantu1999,ma2000oscillators}. For the purposes of this research, let us assume that a vector field $\bsu(\bsx,t)$ (\textit{e.g.} a time-varying fluid velocity field, such as a turbulent flow) can be decomposed as
\begin{equation}
    \label{eq:pod}
    \bsu(\bsx,t) = \bsU(\bsx) + \sum_{i=1}^\infty 
    \boldsymbol{\psi}_i(\bsx)\boldsymbol{\phi}_i(t),
\end{equation}
that is, the sum of a reference state $\bsU$ (oftentimes the time-average of the velocity field), and an orthonormal basis of spatial functions $\{\boldsymbol{\psi}_i(x)\}_{i=1}^\infty$. Each of these spatial modes maximizes the captured energy content according to the following condition
\begin{equation}
    \int_\Omega \mathbb{E}
    [\bsu'(\bsx,t)\bsu'(\bsx',t)]\boldsymbol{\psi}_i(\bsx')d\bsx'=\lambda_i \boldsymbol{\psi}_i(\bsx),
\end{equation}
where $\mathbb{E}$ represents the expected value, and $\bsu' = \bsu - \bsU$ is the mean-subtracted velocity field. 
The temporal coefficients $\boldsymbol{\phi}_i(t)$ corresponding to each of the spatial modes can be computed  from the integral
\begin{equation}
\label{eq:coefficients}
    \boldsymbol{\phi}_i(t) = \int_\Omega \bsu(\bsx,t) \boldsymbol{\psi}_i(\bsx,t) d\bsx.
\end{equation}
Notice that while these definitions 
yield POD coefficients $\boldsymbol{\phi}_i(t)$ that are uncorrelated by design, there can still be causal relationships between these coefficients. In this work, this formulation is adapted in order to manipulate finite-dimensional data. This can be formulated as performing a singular value decomposition (SVD) on a data matrix $\bsY$, where each column represents one snapshot of the flow field (\textit{e.g.}~the snapshot shown in Figure~\ref{fig:2snapshots}). 

Here we only consider streamwise-constant structures over the length of the short duct, and therefore we take the average in the $x$-direction. The square duct has a total of 8 geometric (\textit{i.e.}~reflection and rotation) symmetries, which we can enforce to increase the total size of our dataset. 

The data matrix $\bsY$ has dimensions $[3N_yN_z \times 8 N_t]$, where the factor of 3 accounts for all velocity components, and the factor of 8 accounts for all transformations that respect the geometric symmetries. The SVD of the discretized operator $\bsY$ is now written as 
\begin{equation}
    \label{eq:svd}
    \bsY = \boldsymbol{\Psi} \Sigma \boldsymbol{\Phi}^*,
\end{equation}
where the right-hand side represents the product of discrete operators. Each of the entries $\sigma_i$ in the diagonal matrix $\boldsymbol{\Sigma}$ is related to the energy content of the spatial mode $\boldsymbol{\psi}_i$ (the $i$-th column of $\boldsymbol{\psi}$) associated with a temporal coefficient $\boldsymbol{\phi}_i$ (the $i$-th column of $\boldsymbol{\Phi}$). The POD modes for the duct configuration will be shown and discussed in Section \ref{sec:results_pod} (Figures~\ref{fig:PODmodes}--\ref{fig:PODmodes_split}).

\section{Causality analysis}
\label{sec:causality}
Identifying and characterizing the mechanisms that exist within dynamically-complex fluid flows can be an arduous task. However, the study of causal relationships between some of the identified mechanisms or features can potentially unveil relevant features or processes that would be otherwise difficult to trace. We ultimately seek to identify relationships of causality among the streamwise-constant structures that play a part in sustaining and driving turbulence within a rectangular/square straight duct. These relationships could potentially provide further insight concerning the energy flux between scales and velocity components, and also could inform us about the role of the secondary motions. 

In this work, we consider a multivariate formulation of Granger causality analysis (MVGC) \cite{barnett2014mvgc,barnett2015granger} to study the pairwise causal relationships between the temporal coefficients ($\boldsymbol{\phi}_i(t)$ in Equation~\eqref{eq:svd}) of the streamwise-constant coherent structures within the minimal duct. This framework provides a quantitative estimation of the amount of information that is transferred from one variable (captured by its time-series) to another, while accounting for the effect of knowledge of all other variables. 
 We first discuss this for standard (linear) Granger causality, before describing a nonlinear extension to this method.
 
\subsection{Linear multivariate Granger causality analysis}
Granger causality analysis is a statistics-based test first developed by Granger~\cite{granger1969} to measure the causal relationship between econometric time series or variables \cite{berchtold2002,gaussiansignals2020}. Here, the concept of causality is studied through the lens of predictability: the causal relationship between two variables $\boldsymbol{\phi}_i$ and $\boldsymbol{\phi}_j$ measures how much the previous knowledge of $\boldsymbol{\phi}_j$ improves the prediction of $\boldsymbol{\phi}_i$. 
Let us formalize this concept and consider a set $\boldsymbol{\phi}(t)=[\boldsymbol{\phi}_1(t),\boldsymbol{\phi}_2(t),...,\boldsymbol{\phi}_n(t)]$ of temporal signals (in this work they will coincide with the temporal coefficients of the POD modes). In this framework, the prediction  of the current state of a given signal $\boldsymbol{\phi}_i(t)$ at a time $t$ can be written as
\begin{equation}
\label{eq:granger_with}
    \boldsymbol{\phi}_i(t) = \sum_{r=1}^{p} A^{(r)}_{i,i}\boldsymbol{\phi}_i^{(r)}+\sum_{r=1}^p A_{i,j}^{(r)}\boldsymbol{\phi}_j^{(r)}+\sum_{\substack{k=1 \\ k \notin\{i,j\}}}^n\sum_{r=1}^p A_{i,k}^{(r)}\boldsymbol{\phi}_k^{(r)}+\epsilon_{i,j}(t),
\end{equation}
where here we use the shorthand $\boldsymbol{\phi}_i^{(r)} = \boldsymbol{\phi}_i(t-r\Delta t)$ to denote the previous-time measurements, where we use a total of $p$ prior timesteps in the model for the current state. Note that this parameter $p$ will be referred to in the results section as the time lag. The sum over $k$ involves all indices from 1 to $n$ excluding $i$ and $j$, which makes the causal analysis between $i$ and $j$ conditional on the effect of all other variables. This definition 
prescribes a linear relationship between a  
temporal signal $\boldsymbol{\phi}_i$, and past values of this and all other variables (across a time horizon $p\Delta t$). In this work, the coefficients $A$ in the linear model are fitted from data. 
The term $\epsilon_{i,j}(t)$ denotes the error in this fitted linear model.
Notice that the contribution of the variable $\boldsymbol{\phi}_j$, where $j\neq i$, has been isolated on the second term of the right-hand side, and if we were to neglect it, we could write a modified prediction of $\boldsymbol{\phi}_i(t)$ as
\begin{equation}
\label{eq:granger_without}
    \boldsymbol{\phi}_i(t) = \sum_{r=1}^p A_{i,i}^{(r)}\boldsymbol{\phi}_i^{(r)}+\sum_{\substack{k=1 \\ k \notin\{i,j\}}}^n\sum_{r=1}^p A_{i,k}^{(r)}\boldsymbol{\phi}_k^{(r)}+\epsilon_{i,\bcancel{j}}(t).
\end{equation}
Note that the fitted coefficients $A$ in equation \ref{eq:granger_without} will be different to those in equation \ref{eq:granger_with}, owing to having fewer terms on the right hand side. 
These two expressions allow us to infer the pairwise causal relationship between the temporal signals $\boldsymbol{\phi}_i$ and $\boldsymbol{\phi}_j$ by looking at the difference between the size of the residuals $\epsilon_{\i,j}(t)$ and $\epsilon_{i,\bcancel{j}}(t)$. Hence, in this framework, causality between two variables $\boldsymbol{\phi}_i$ and $\boldsymbol{\phi}_j$ is measured as the difference between the errors in the predictions of $\boldsymbol{\phi}_i$ including and excluding previous knowledge of $\boldsymbol{\phi}_j$.%

One of the most relevant implications of this method is the fact that causality can only be found in the history of a time series, and therefore can be sensitive to the observed time window of the temporal signals. Moreover, this formulation restricts the causality analysis to linear mechanisms, potentially missing  relevant nonlinear dynamics. It can be additionally shown that Granger causality is equivalent to transfer entropy (which has been used for causality analysis of fluids problems in several previous works \cite{lozanoduran2022causality,martinez2022causality,srivastava2021causality}) in the case of Gaussian variables \cite{barnett2009granger}.

\subsection{Nonlinear multivariate Granger causality analysis}
One significant drawback of Granger causality analysis is that it is based on a  linear modeling assumption. Consequently, the residuals $\epsilon_{i,j}$ and $\epsilon_{i,\bcancel{j}}$ in Equations~\eqref{eq:granger_with} and \eqref{eq:granger_without} likely include the contribution of nonlinear terms involving the variables $\boldsymbol{\phi}_i$, $\boldsymbol{\phi}_j$, and $\boldsymbol{\phi}_k$. To make this approach viable for nonlinear systems, we propose an extension of the original formulation of multivariate Granger causality that allows for the presence of nonlinear terms. In particular, the proposed formulation can be extended to systems with quadratic nonlinear terms  while largely maintaining the benefits and simplicity of its linear counterpart. Hence, the causal relationships are still quantified as prediction errors, but after accounting for the presence of selected nonlinear terms. Hence, we can write the general form of the second-order prediction of the current state of a time series $\boldsymbol{\phi}_i(t)$ as a function of the history of the variables $\boldsymbol{\phi}_i$, $\boldsymbol{\phi}_j$ and $\{\boldsymbol{\phi}_k\}_{k \notin \{i,j\}}$ 
\begin{equation}
\label{eq:nonlingranger_with}
\begin{split}
    \boldsymbol{\phi}_i(t) & = \sum_{r=1}^p A_{i,i}^{(r)}\boldsymbol{\phi}_i^{(r)}+\sum_{r=1}^p A_{i,j}^{(r)}\boldsymbol{\phi}_j^{(r)}+\sum_{\substack{k=1 \\ k \notin\{i,j\}}}^n\sum_{r=1}^p A_{i,k}^{(r)}\boldsymbol{\phi}_k^{(r)}+\sum_{r=1}^p Q_{i,i,i}^{(r)}(\boldsymbol{\phi}_i \boldsymbol{\phi}_i)^{(r)}+\\
    &\sum_{r=1}^p Q_{i,i,j}^{(r)}(\boldsymbol{\phi}_i \boldsymbol{\phi}_j)^{(r)}+\sum_{\substack{k=1 \\ k \notin\{i,j\}}}^n\sum_{r=1}^p Q_{i,i,k}^{(r)}(\boldsymbol{\phi}_i  \boldsymbol{\phi}_k)^{(r)}+\sum_{r=1}^p Q_{i,j,j}^{(r)}(\boldsymbol{\phi}_j  \boldsymbol{\phi}_j)^{(r)}+\\
    &\sum_{\substack{k=1 \\ k \notin\{i,j\}}}^n\sum_{r=1}^p Q_{i,j,k}^{(r)}(\boldsymbol{\phi}_j \boldsymbol{\phi}_k)^{(r)}+\sum_{\substack{k'=1 \\ k' \notin\{i,j\}}}^n\sum_{\substack{k=1 \\ k \notin\{i,j\}}}^n\sum_{r=1}^p Q_{i,k',k}^{(r)}(\boldsymbol{\phi}_{k'}  \boldsymbol{\phi}_k)^{(r)}+ \epsilon_{i,j}'(t).
\end{split}
\end{equation}
Here, the $A$-terms capture the linear interactions and the $Q$-terms model the quadratic interactions. We can similarly write a prediction of the current state $\boldsymbol{\phi}_i(t)$ while excluding all terms that require knowledge of the history of the variable $\boldsymbol{\phi}_j$ as follows
\begin{equation}
\label{eq:nonlingranger_without}
\begin{split}
    \boldsymbol{\phi}_i(t) & = \sum_{r=1}^p A_{i,i}^{(r)}\boldsymbol{\phi}_i^{(r)}+\sum_{\substack{k=1 \\ k \notin\{i,j\}}}^n\sum_{r=1}^p A_{i,k}^{(r)}\boldsymbol{\phi}_k^{(r)}+\sum_{r=1}^p Q_{i,i,i}^{(r)}(\boldsymbol{\phi}_i  \boldsymbol{\phi}_i)^{(r)}+\\
    &\sum_{\substack{k=1 \\ k \notin\{i,j\}}}^n\sum_{r=1}^p Q_{i,i,k}^{(r)}(\boldsymbol{\phi}_i \boldsymbol{\phi}_k)^{(r)}+\sum_{\substack{k'=1 \\ k' \notin\{i,j\}}}^n\sum_{\substack{k=1 \\ k \notin\{i,j\}}}^n\sum_{r=1}^p Q_{i,k',k}^{(r)}(\boldsymbol{\phi}_{k'}  \boldsymbol{\phi}_k)^{(r)}+ \epsilon_{i,\bcancel j}'(t),
\end{split}
\end{equation}
The causal relationship between $\boldsymbol{\phi}_i$ and $\boldsymbol{\phi}_j$ is then again estimated from the difference in magnitude between the residuals of these models, $\epsilon_{i,j}'(t)$ and $\epsilon_{i,\bcancel j}'(t)$. While this methodology has been formulated in terms of the POD coefficients, in Section~\ref{sec:results} we will also consider the case where the variables in the identified models are instead the squares of the POD coefficients, $\boldsymbol{\phi}_i^2$. The dynamics of these variables may indeed need higher order nonlinearities to be fully modeled, though such analysis is beyond the scope of the present investigation. Finding a low-dimensional nonlinear model from data can be performed in a number of ways. Here, this is done by fitting coefficients to data, though it would also be possible to obtain such models by Galerkin projection of the governing equations onto POD modes \cite{holmes1996}. This data-driven identification of a nonlinear model also shares similarity with the sparse identification of nonlinear dynamics (SINDy) methodology, which incorporates sparsity-promoting methods to reduce the number of nonzero coefficients
 \cite{sindy2016brunton}.

\section{Results}
\label{sec:results}
Here, we discuss the main findings of this research. Section~\ref{sec:moehlis} first present preliminary results that validate the formulation of the nonlinear Granger causality analysis on the nine-equation model for turbulent shear flows introduced by Moehlis et al.~\cite{moehlis2004}. This is followed by a discussion of the observed POD modes and the consequent Granger causality analysis using the corresponding temporal coefficients in Section~\ref{sec:results_pod}.

\subsection{Linear and nonlinear Granger causality analysis on 
a low-order turbulence model
}
\label{sec:moehlis}
\begin{figure}[ht!]
\vspace*{-0.cm}
\centering {
{\hspace*{0.6cm}\includegraphics[width=0.5\textwidth]{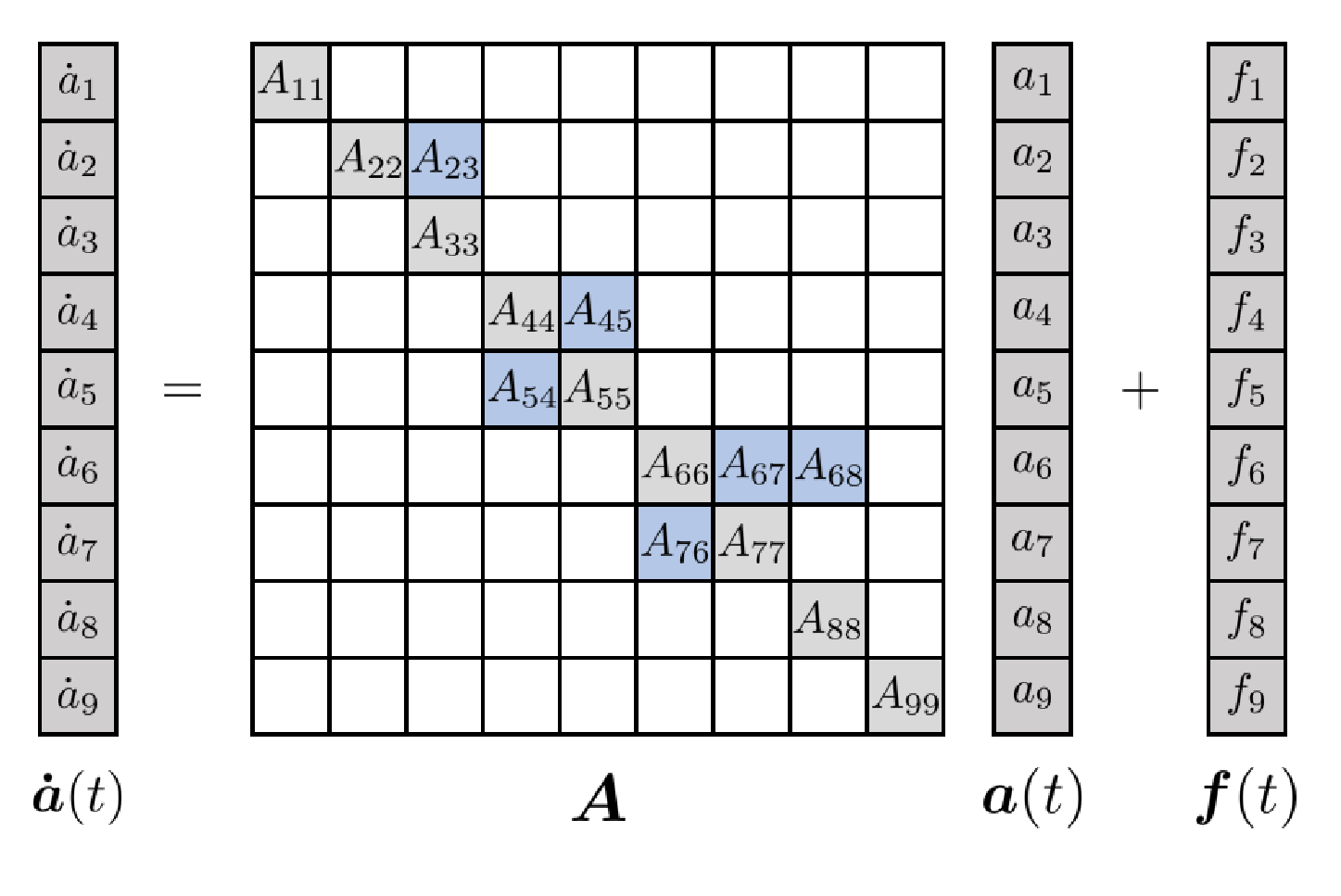}} }
\vspace*{-0.15cm}
\caption{Schematic of the linearized form of the temporal coefficients $a_j(t)$ of the nine equation model \cite{moehlis2004}. The elements $A_{ij}$ in the linear operator $\bsA$ that are shaded in grey represent the linear viscous terms,  those in blue represent the linear elements arising due to the mean flow (\textit{i.e.}~analogous to mean shear), and the white elements correspond to zero entries.}
\vspace*{-0.2cm}
\label{fig:moehlisDiagram}
\end{figure}

In order to assess the validity and capabilities of the proposed nonlinear Granger causality analysis, we first test its performance on the low-dimensional model of wall-bounded turbulent flows introduced by Moehlis et al.~\cite{moehlis2004}. This model is a set of 9 coupled differential equations, featuring quadratic nonlinearities. The variables correspond to features of wall-bounded turbulence, including the mean streamwise velocity, spanwise flow, streamwise streaks, 
streamwise vortices and so forth. The temporal coefficients $a_j(t)$ associated to each of the entries in the low-dimensional model have the closed solutions in the following form 
{\footnotesize
\begin{align*}
    &\dot{a}_1(t) = \frac{\beta^2}{Re} (1-a_1) - \sqrt{\frac{3}{2}} \beta \gamma \left(\frac{a_6 a_8}{\kappa_{\alpha\beta\gamma}}- \frac{a_2 a_3}{\kappa_{\beta\gamma}}\right), \\   
    &\dot{a}_2(t) = -\frac{a_2}{Re}\left(\frac{4\beta^2}{3}+\gamma^2 \right)+\frac{\gamma^2}{\kappa_{\alpha\gamma}}\left(\frac{5\sqrt{2}}{3\sqrt{3}}a_4a_6-\frac{1}{\sqrt{6}}a_5a_7 \right)  
    -\frac{\alpha\beta\gamma}{\sqrt{6} \kappa_{\alpha\gamma}\kappa_{\alpha\beta\gamma}}a_5a_8-\sqrt{\frac{3}{2}}\frac{\beta\gamma}{\kappa_{\beta\gamma}}(a_1a_3-a_3a_9),\\
    &\dot{a}_3(t) = -\frac{\beta^2+\gamma^2}{Re}a_3+\frac{2}{\sqrt{6}}\frac{\alpha\beta\gamma}{\kappa_{\alpha\gamma}\kappa_{\beta\gamma}}(a_4 a_7+a_5 a_6)+\frac{\beta^2(3\alpha^2+\gamma^2)-3\gamma^2(\alpha^2+\gamma^2)}{\sqrt{6}\kappa_{\alpha\gamma} \kappa_{\beta\gamma}\kappa_{\alpha\beta\gamma}}a_4 a_8,\\
    &\dot{a}_4(t) =  -\frac{3\alpha^2+4\beta^2}{3 Re}a_4 - \frac{\alpha}{\sqrt{6}}a_1 a_5 - \frac{10}{3\sqrt{6}}\frac{\alpha^2}{\kappa_{\alpha\gamma}}a_2 a_6  -\sqrt{\frac{3}{2}}\frac{\alpha\beta\gamma}{\kappa_{\alpha\gamma}\kappa_{\beta\gamma}}a_3a_7 -\sqrt{\frac{3}{2}}\frac{\alpha^2\beta^2}{\kappa_{\alpha\gamma}\kappa_{\beta\gamma}\kappa_{\alpha\beta\gamma}}a_3a_8 -\frac{\alpha}{\sqrt{6}}a_5a_9   ,
       \end{align*}
\begin{align*}
   & \dot{a}_5(t) = -\frac{\alpha^2+\beta^2}{Re}a_5 +\frac{\alpha}{\sqrt{6}}a_1a_4+\frac{\alpha^2}{\sqrt{6}\kappa_{\alpha\gamma}}a_2a_7   -\frac{\alpha\beta\gamma}{\sqrt{6}\kappa_{\alpha\gamma}\kappa_{\alpha\beta\gamma}}a_2a_8+\frac{\alpha}{\sqrt{6}}a_4a_9+\frac{2}{\sqrt{6}}\frac{\alpha\beta\gamma}{\kappa_{\alpha\gamma}\kappa_{\beta\gamma}}a_3a_6  ,\\
    & \dot{a}_6(t) = -\frac{3\alpha^2+4\beta^2+3\gamma^2}{3Re}a_6+\frac{\alpha}{\sqrt{6}}a_1a_7+\sqrt{\frac{3}{2}}\frac{\beta\gamma}{\kappa_{\alpha\beta\gamma}}(a_1a_8+a_8a_9)+\frac{10}{3\sqrt{6}}\frac{\alpha^2-\gamma^2}{\kappa_{\alpha\gamma}}a_2 a_4-2\sqrt{\frac{2}{3}}\frac{\alpha\beta\gamma}{\kappa_{\alpha\gamma}\kappa_{\beta\gamma}}a_3a_5+\frac{\alpha}{\sqrt{6}}a_7a_9 ,\\
    & \dot{a}_7(t) = -\frac{\alpha^2+\beta^2+\gamma^2}{Re}a_7-\frac{\alpha}{\sqrt{6}}(a_1a_6+a_6a_9)+\frac{1}{\sqrt{6}}\frac{\gamma^2-\alpha^2}{\kappa_{\alpha\gamma}}a_2a_5+\frac{1}{\sqrt{6}}\frac{\alpha\beta\gamma}{\kappa_{\alpha\gamma}\kappa_{\beta\gamma}}a_3a_4, \\
    & \dot{a}_8(t) = -\frac{\alpha^2+\beta^2+\gamma^2}{Re}a_8+\frac{2}{\sqrt{6}}\frac{\alpha\beta\gamma}{\kappa_{\alpha\gamma}\kappa_{\alpha\beta\gamma}}a_2a_5+\frac{\gamma^2(3\alpha^2-\beta^2+3\gamma^2)}{\sqrt{6}\kappa_{\alpha\gamma}\kappa_{\beta\gamma}\kappa_{\alpha\beta\gamma}}a_3a_4, \\
    & \dot{a}_9(t) = -\frac{9\beta^2}{Re}a_9+\sqrt{\frac{3}{2}}\frac{\beta\gamma}{\kappa_{\beta\gamma}}a_2a_3-\sqrt{\frac{3}{2}}\frac{\beta\gamma}{\kappa_{\alpha\beta\gamma}}a_6a_8,    
\end{align*}}
 where $\kappa_{\alpha\gamma}=\sqrt{\alpha^2+\gamma^2}$, $\kappa_{\beta\gamma}=\sqrt{\beta^2+\gamma^2}$, $\kappa_{\alpha\beta\gamma}=\sqrt{\alpha^2+\beta^2+\gamma^2}$. Here, $\alpha$ and $\gamma$ represent the streamwise and spanwise wavenumbers, respectively, and $\beta=\pi/2$. This system of nonlinear equations can be linearized about the turbulent mean, resulting in equations of the form $\dot{\bsa}(t)=\bsA \bsa (t)+\bsf(t)$, where $\bsf(t)$ is a forcing to the linearized dynamics that can include the effect of nonlinear and constant terms. 
The mean state of this system is nonzero in the first and ninth states. This leads to quadratic terms involving $a_1$ and $a_9$ giving nonzero off-diagonal entries of $\bsA$, as shown in Figure~\ref{fig:moehlisDiagram}. 

Due to the presence of nonlinear terms in these equations, a causality analysis limited to linear interactions would likely not be sufficient for identifying pertinent causal mechanisms. We use linear and nonlinear Granger causality methods to study the interactions between the nine variables of this model, and the results are produced in the form of causality maps. That is, square diagrams in which each of the entries represents one variable, the horizontal axis indicates the origin of the information flux and the vertical axis represents the destination of this flux (\textit{e.g.}~a large entry in the element (2,3) of the causality map indicates that there is a relevant causal interaction between variables 3 and 2, in which 3  causes 2). See \cite{barnett2014mvgc,barnett2015granger} for more details concerning how such maps are computed from data via identification of the equations described in Section \ref{sec:causality}.

\begin{figure}[ht!]
\vspace*{-0.cm}
\centering {
{\hspace*{-0.2cm}\includegraphics[width=1.\textwidth]{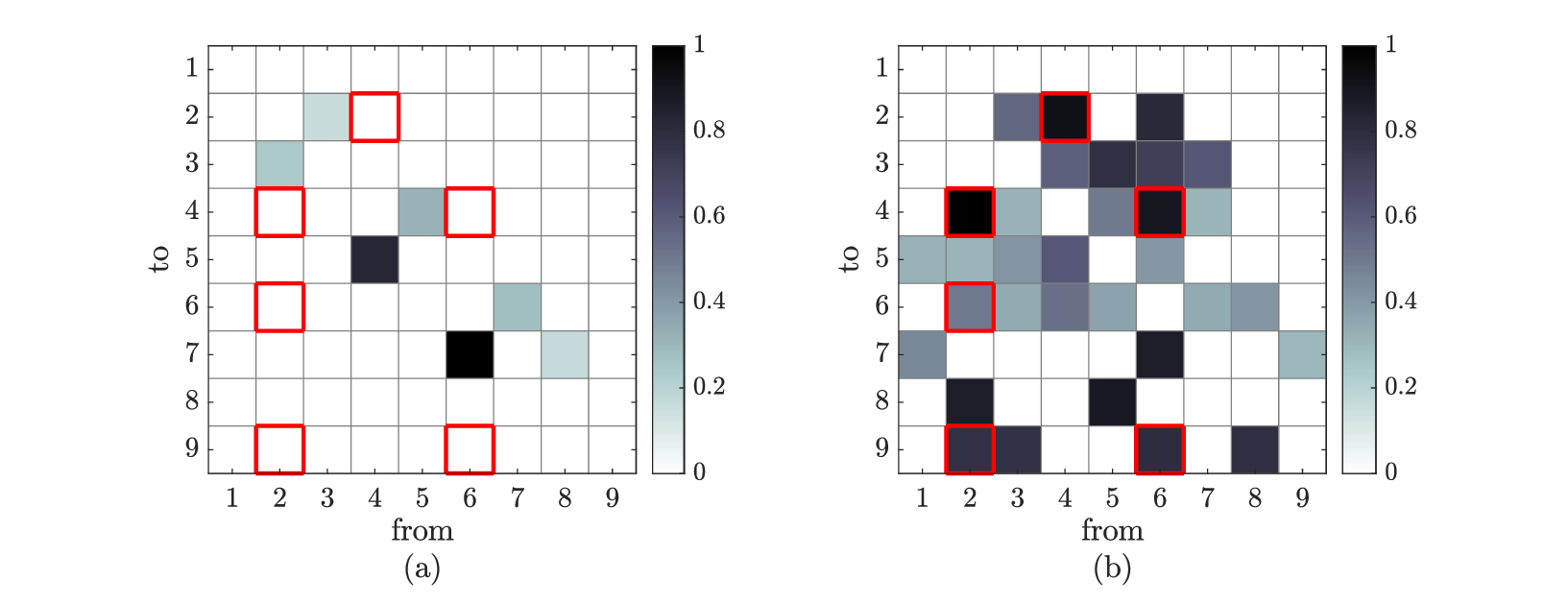}} }
\vspace*{-0.5cm}
\caption{Causality maps produced for the nine-equation model described in Moehlis et al.~\cite{moehlis2004} using (a) linear Granger and (b) nonlinear Granger pairwise multivariate causality analyses. A time lag $p$ corresponding to one time step of size $\Delta t=1$ was considered. The magnitude of the causality maps has been normalized by the largest entry. Entries highlighted in red are the principal causal relationships identified in Mart{\'\i}nez-S{\'a}nchez et al.~\cite{martinez2022causality}.
}
\vspace*{-0.2cm}
\label{fig:resultsMoehlis}
\end{figure}

The causality maps shown in Figure~\ref{fig:resultsMoehlis} are computed considering 600 different trajectories in a time window $0\leq t \leq 4,000$ with $\Delta t=1$, where the first 100 snapshots are discarded. The results described in Mart{\'\i}nez-S{\'a}nchez et al.~\cite{martinez2022causality} are taken as reference, where a transfer entropy causality analysis was performed.
We observe that the results provided by the linear formulation of Granger causality analysis in Figure~\ref{fig:resultsMoehlis}(a) are not sufficiently accurate, mainly due to the fact that most of the nonlinear interactions are missed. Indeed, the causal interactions identified here correspond to several of the nonzero terms in the linearized form of this model (\textit{e.g.}~the terms $A_{76}$ and $A_{54}$ in Figure~\ref{fig:moehlisDiagram}). Note that while the causal mechanisms in the linear Granger analysis (of nonlinear data) do not exactly correlate to all of the nonzero terms in the mean-linearized model, the causality map in figure Figure~\ref{fig:resultsMoehlis}(a) has the same block diagonal structure as the mean-linearized dynamics.

In contrast, the nonlinear formulation of Granger causality analysis is able to identify many additional causal relationships between the variables.  In particular, the six major causal interactions identified using transfer entropy in Mart{\'\i}nez-S{\'a}nchez et al.~\cite{martinez2022causality}, highlighted in red in Figure~\ref{fig:resultsMoehlis}), are also identified as causal mechanisms here. These entries involve some of the nonlinear term (off-diagonal entries) in Figure~\ref{fig:moehlisDiagram}, and represent mechanisms such as: the effect of streamwise streaks and wall-normal vortices modifying the mean flow in entries (9,2) and (9,6), respectively; the influence of streamwise streaks on the wall-normal vortices in the element (6,2); the excitation of the spanwise flow by the wall-normal vortices in entry (4,6); and the interactions between the streamwise streaks and the spanwise flow by the elements (4,2) and (2,4).

The nonlinear Granger method also identifies a number additional significant causal mechanisms not identified by transfer entropy. The reasons for these  differences are the subject of ongoing study. This also highlights an aspect of causality analysis that will repeatedly emerge in the following analysis: the precise causal mechanisms identified can be highly sensitive to the methods used.

\subsection{Causal interactions between streamwise-constant coherent structures in turbulent duct flow}
\label{sec:results_pod}
Here we first present the spatial POD modes $\boldsymbol{\psi}_j$ (computed according to Equation~\eqref{eq:svd}), extracted from the 4000$\times$8 (accounting for all eight geometric rotations) snapshots of available data $\bsY$ within the minimal duct. Notice that each of these snapshots has been averaged in the streamwise direction, since the goal of this study is to focus on streamwise-elongated structures. Due to the relations of symmetry within the $(y,z)$ plane, the decomposition is restricted to the lower-left corner of the duct, in order to isolate the relevant coherent structures without accounting for the symmetrical or anti-symmetrical counterparts that may arise in other regions of the cross-section (It would also be possible to instead study only one eighth of the duct using similar symmetry arguments). 

\begin{figure}[ht!]
\vspace*{-0.2cm}
\centering {
{\hspace*{-1.4cm}\includegraphics[width= 1.18\textwidth]{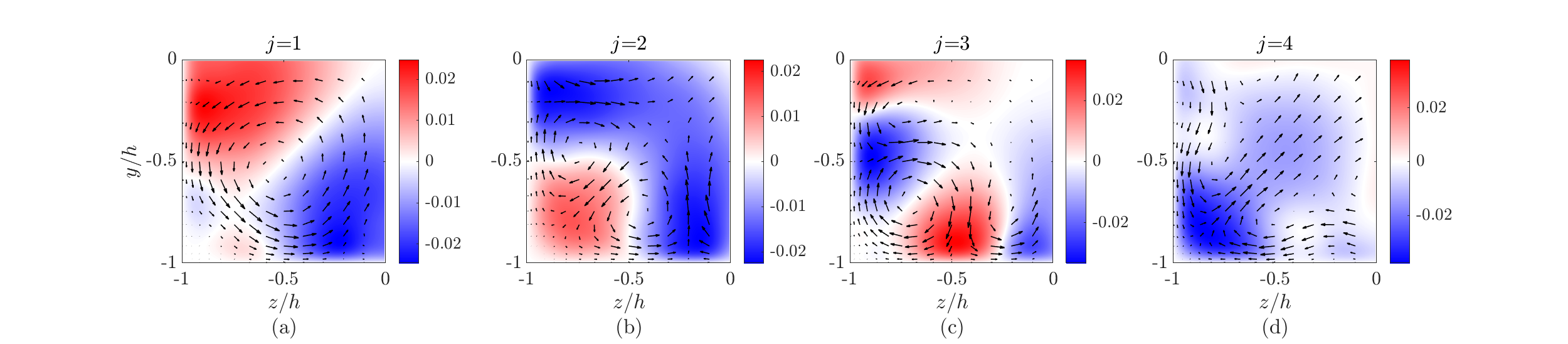}} }
\vspace*{-0.3cm}
\caption{Spatial modes $\boldsymbol{\psi}_j$ with $j$=\{1,2,3,4\} extracted from the full snapshots in the lower-left corner of the minimal duct, with $Re_\tau=180$ and $A=1$. The filled contour plots represent the streamwise component, and the secondary velocity components are depicted as vector fields.
}
\vspace*{-0cm}
\label{fig:PODmodes}
\end{figure}

The first set of results is produced via POD of the full set of snapshots formed by all three velocity components simultaneously, which yields the {full} spatial POD modes depicted in Figure~\ref{fig:PODmodes}. Each of these modes contains regions of high and low (negative) streamwise velocity fluctuation, indicative of streaks located in the duct corner (as in modes 2 and 4), or of opposite sign on adjacent walls (modes 1 and 3). Note that modes 1 and 3 are antisymmetric about the bisector of the corner, while modes 2 and 4 are symmetric. Modes 1 and 3 have a secondary ($v$ and $w$) flow characterized by a single streamwise vortex between the fast and slow streaks, while modes 2 and 4 have a pair of counter-rotating vortices. Modes 2 and 4 therefore resemble the structure of the secondary mean flow (Figure ~\ref{fig:turbulentMean}~(b)). We emphasize that the mean flow has been subtracted before computing POD, so modes 2 and 4 would correspond to transient fluctuations in the nature and size of the structures giving rise to the secondary mean.

\begin{figure}[ht!]
\vspace*{-0.0cm}
\centering {
{\hspace*{-2cm}\includegraphics[width= 1.25\textwidth]{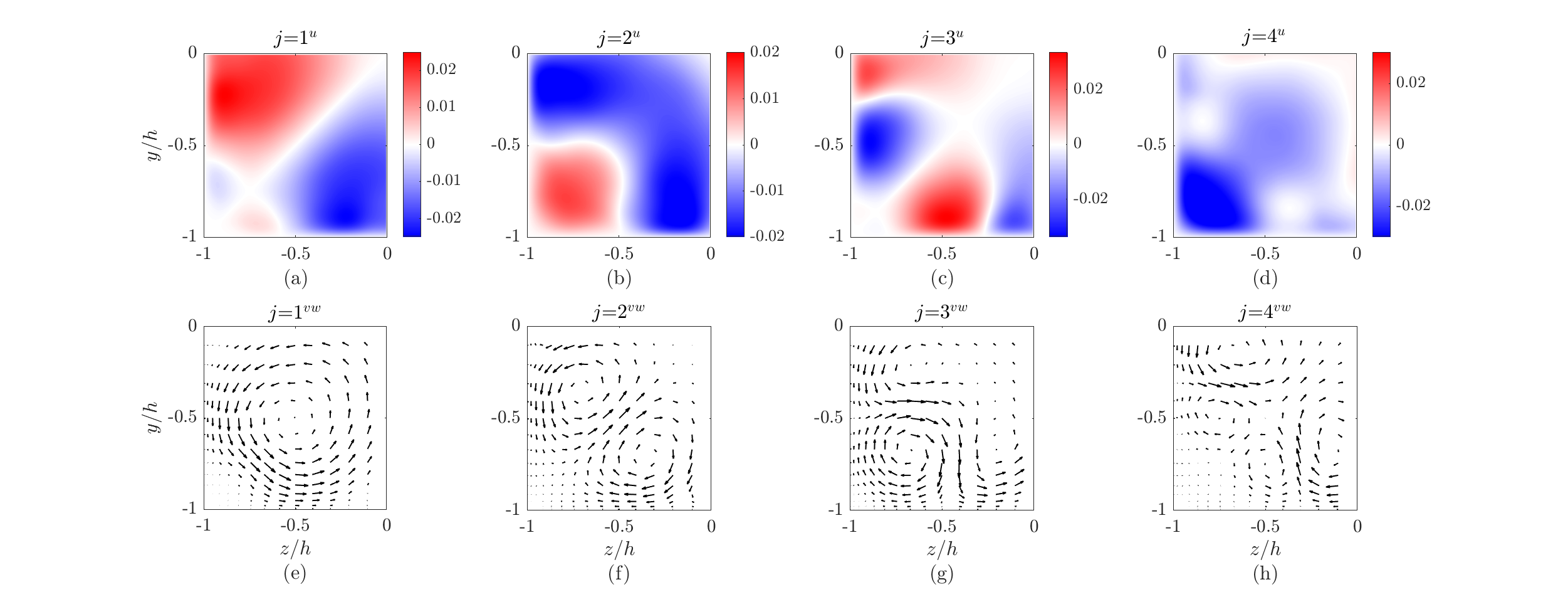}} }
\vspace*{-0.5cm}
\caption{Spatial modes $\boldsymbol{\psi}_j$ with $j$=\{1,2,3,4\}$^u$(a)-(d) and $j$=\{1,2,3,4\}$^{vw}$(e)-(h) extracted from the split snapshots, in the lower-left corner of the minimal duct, with $Re_\tau=180$ and $A=1$. The filled contour plots represent the streamwise component, and the secondary velocity components are depicted as vector fields.
}
\vspace*{-0.cm}
\label{fig:PODmodes_split}
\end{figure}

The second set of results is obtained via POD of the decoupled snapshots, that is, the decomposition is performed on the streamwise component $u$ and the $(v,w)$ components of the snapshots separately. This is done to enable causal analysis between the streamwise and secondary velocity components. 
Spatial modes $\boldsymbol{\psi}_j$ with $j$=\{1,2,3,4\}$^u$ and $j$=\{1,2,3,4\}$^{vw}$ from each of the decoupled sets are shown in Figure~\ref{fig:PODmodes_split}. Notice that the streamwise  velocity components of these modes are similar to those shown in each of the four modes in Figure~\ref{fig:PODmodes}. The secondary components are also generally similar, particularly for modes 1 and 3. For mode 2 (Figure~\ref{fig:PODmodes_split}(f)) the part of streamwise vorticies is larger than in Figure~\ref{fig:PODmodes}(b). 
For mode 4, the pattern depicted in Figure~\ref{fig:PODmodes_split}(h) differs from the secondary components of the mode in Figure~\ref{fig:PODmodes}(d). This highlights the fact that when POD modes are computed using all components, the identified structures are driven by the energetically-dominant streamwise velocity component. 
Also note that since the computations are performed separately, now there can also be correlation between the coefficients of the streamwise and secondary velocity modes. 

\begin{figure}[ht!]
\vspace*{-0.0cm}
\centering {
{\hspace*{-0cm}\includegraphics[width= 1\textwidth]{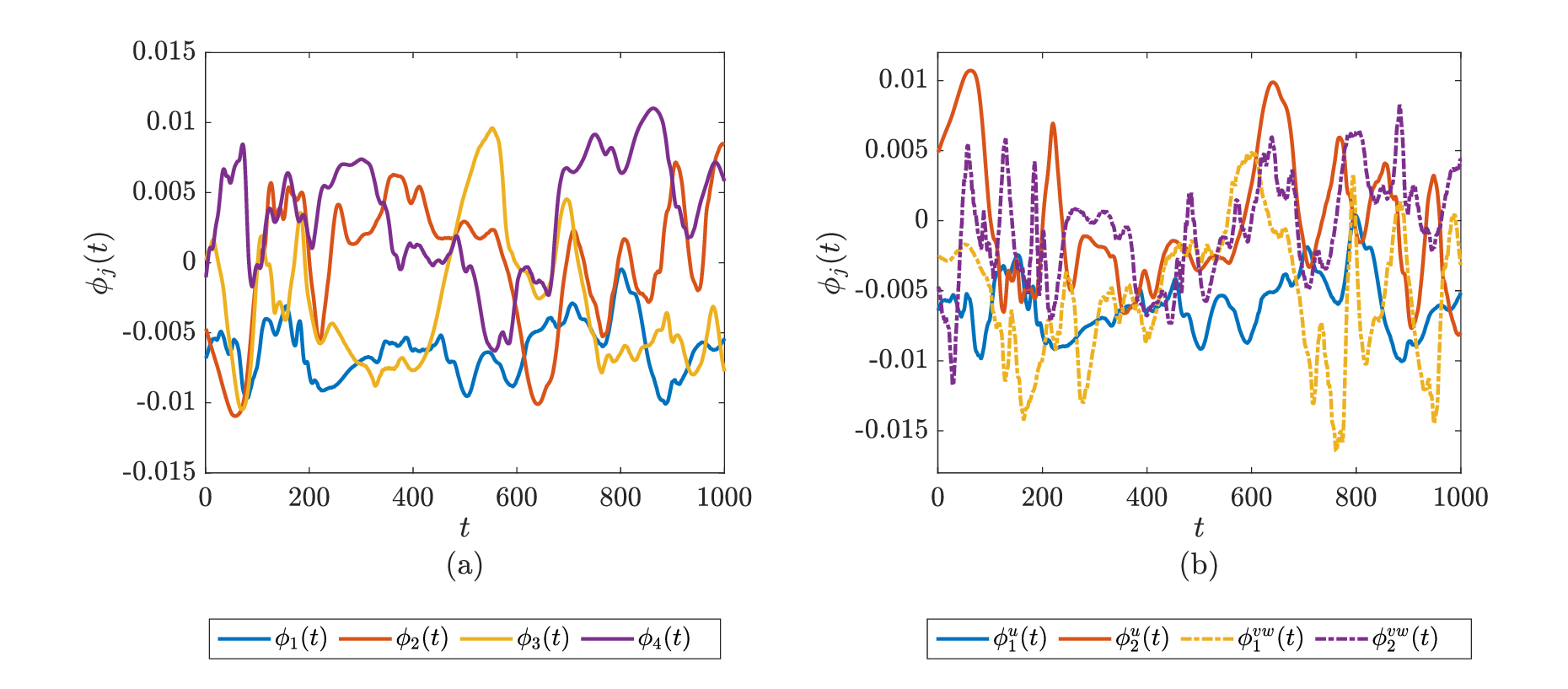}} }
\vspace*{-0.2cm}
\caption{Temporal coefficients $\boldsymbol{\phi}_j(t)$ associated with POD modes $\boldsymbol{\psi}_j$ with $j$=\{1,2,3,4\} in Figure~\ref{fig:PODmodes} and POD modes $\boldsymbol{\psi}_j$ with $j$=\{1,2\}$^u$ and $j$=\{1,2\}$^{vw}$ in Figure \ref{fig:PODmodes_split}, over a time window $0\leq t \leq 1,000$ with $\Delta t=1$.
}
\vspace*{-0.2cm}
\label{fig:coefficients}
\end{figure}

Figure \ref{fig:coefficients} shows a portion of several of the time series of the temporal signals associated with modes shown in Figures \ref{fig:PODmodes}--\ref{fig:PODmodes_split}. Note that while this correlation between the streamwise and secondary velocity modes is perhaps observable in Figure \ref{fig:coefficients}(b), this correlation is by no means perfect, indicating some independence between the streamwise and secondary components of the modes shown in Figure \ref{fig:PODmodes}. These figures also show that coherent features can persist in the flow over long time horizons, for example the negative value of the coefficient of the first mode in Figure \ref{fig:coefficients}(a) is maintained over the entire time horizon shown in the figures.

\begin{figure}[ht!]
\vspace*{-0.1cm}
\centering {
{\hspace*{-2.0cm}\includegraphics[width= 1.25\textwidth]{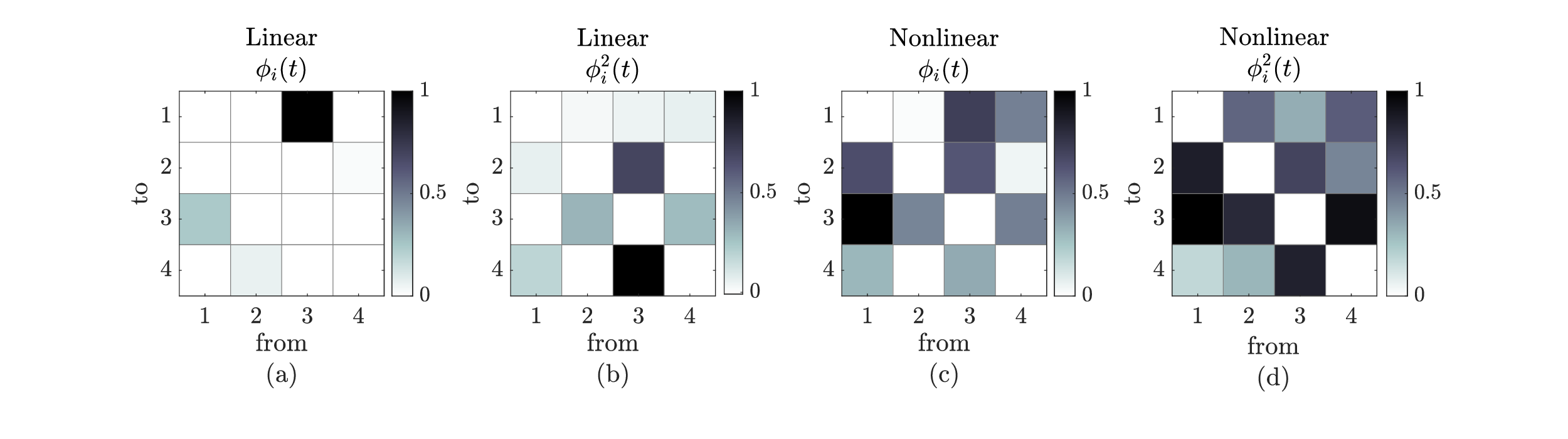}} }
\vspace*{-0.8cm}
\caption{Causality maps produced for the full temporal coefficients $\boldsymbol{\phi}_j(t)$ associated with the spatial modes in Figures~\ref{fig:PODmodes}, as well as the square of the given coefficients, via linear and nonlinear Granger pairwise multivariate causality analyses. The chosen time lag is $p=1$ with $\Delta t$=1 and the entries have been normalized by the largest entry.
}
\vspace*{0.2cm}
\label{fig:causality_full}
\end{figure}

The causality maps generated using the temporal coefficients $\boldsymbol{\phi}_j$ associated to the full modes in Figure~\ref{fig:PODmodes} are shown in Figure~\ref{fig:causality_full}. These causality analyses were performed using both the coefficients extracted from the POD $\boldsymbol{\phi}_j(t)$ and the square of these temporal signals $\boldsymbol{\phi}_j^2(t)$. The latter analysis was performed to see if there were causality mechanisms that are independent of the sign of the cause or effect variable. This may be the case for relationships between symmetric and antisymmetric modes (e.g.~a symmetric mode may have equal chance of causing a positive or negative change in the coefficient of an antisymmetric mode, which could prevent detection if using the coefficients directly). 
The square of the temporal coefficients also represents a measure of the kinetic energy of these coherent structures, hence the identified causal relations could also provide insight concerning the energy transfer mechanisms that drive this system. The chosen time lag for the causality analyses presented here is $p=1$, and different values were not found to have a significant effect on the observed results. 

The linear formulation of Granger causality analysis in Figure \ref{fig:causality_full}(a) highlights one main causal relationship from mode 3 to 1, representing the movement of near-wall streaks and a streamwise vortex along a sidewall away from the corner. There is also a weaker causal analysis in the opposite direction between these two modes. Using the square of the POD coefficients, $\boldsymbol{\phi}^2_j$, as inputs to the linear Granger analysis identifies entirely different causal mechanisms, as shown in Figure \ref{fig:causality_full}(b). In particular, this analysis identifies causal relationships between symmetric and antisymmetric modes, with the the strongest causality identified as mode 3 influencing modes 2 and 4. This mechanism seemingly corresponds to one of the two streaks on adjacent sides of the corner moving into the corner itself, with the streamwise vortex similarly translating. The other causal mechanisms identified in Figure \ref{fig:causality_full}(b) follow similar mechanisms, and the reverse behavior whereby a  streak moves out of the corner along one of the adjacent walls. As mentioned previously, we postulate that the reason why this mechanism is entirely missing from Figure \ref{fig:causality_full}(a) is because of the interaction between symmetric and antisymmetric modes, where here due to the symmetry of the geometry it is equally likely for a corner streak/vortex to move towards either of the adjacent sides. This would ``cancel out" causal relationships when using the coefficients as variables (which can be positive or negative), but is clearly identified when using their energy content, which is agnostic to the sign of the coefficients.

The nonlinear Granger analysis of the full POD mode coefficients shown in Figure  \ref{fig:causality_full}(c) extracts the same mechanisms identified from the linear Granger analyses of both the POD coefficients and their squared values discussed above (Figures \ref{fig:causality_full}(a)-(b)). This is perhaps expected, since the form of the nonlinear model includes both linear and quadratic terms. For completeness, we also show the results of performing nonlinear Granger analysis on the $\boldsymbol{\phi}^2_j$ terms in Figure \ref{fig:causality_full}(d). This identifies an even broader set of causal relationships, though interpretation of these results must be treated with caution, as the evolution of these energy contents would involve cubic nonlinear terms, which are not included in the nonlinear model.

\begin{figure}[ht!]
\vspace*{-0.4cm}
\centering {
{\hspace*{0cm}\includegraphics[width= 0.85\textwidth]{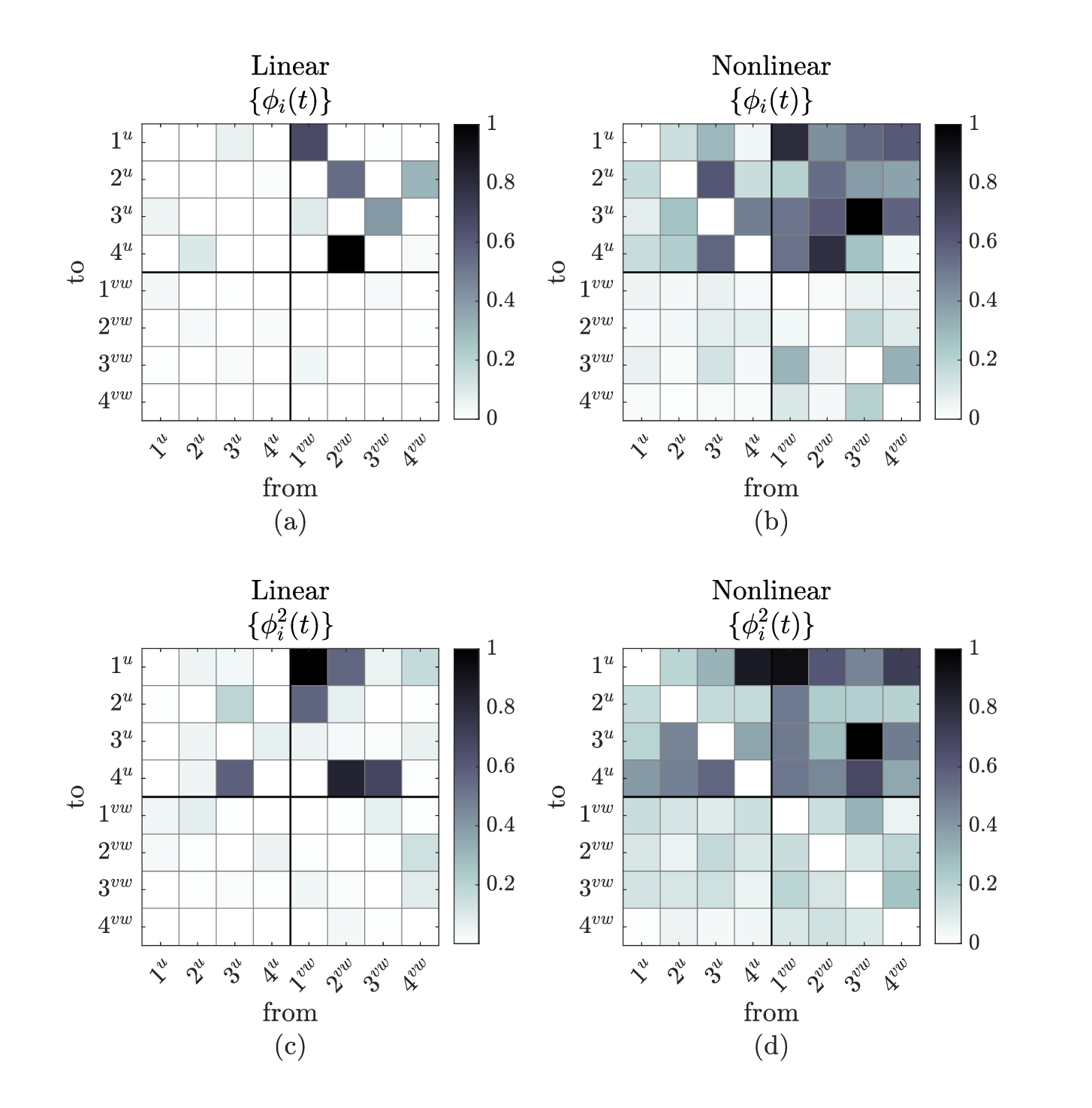}} }
\vspace*{-0.4cm}
\caption{Causality maps produced for the split temporal coefficients $\{\boldsymbol{\phi}_j(t)\}_{split}$ associated with the spatial modes in Figures~\ref{fig:PODmodes_split}, as well as the square of the given coefficients, via linear and nonlinear Granger pairwise multivariate causality analyses. The chosen time lag is $p=1$ with $\Delta t$=1 and the entries have been normalized by the largest entry.
}
\vspace*{-0.0cm}
\label{fig:causality_split}
\end{figure}

The causality maps in Figure \ref{fig:causality_split} correspond to the causality analysis performed on the temporal signals $\boldsymbol{\phi}_j$ and $\boldsymbol{\phi}^2_j$ associated with the split spatial modes shown in Figure~\ref{fig:PODmodes_split}. The temporal coefficients were computed separately in this case to possibly infer relations of causality between velocity components within each mode. In all cases, the upper-right quadrant dominates the causal behavior, emphasizing on the role of the secondary velocity components in driving the formation and dynamics of near-wall and near-corner streamwise streaks. This behavior is likely closely related with the well-known lift-up mechanism \cite{ellingsen1975stability,landahl1980note,brandt2014lift} describing how perturbations in wall-normal velocity can lead to amplified streamwise fluctuations. The causal mechanisms identified by linear Granger causality analysis on the POD coefficients is largely limited to those components corresponding to the same mode for the combined POD, except for the entries corresponding to $2^{vw} \to 4^u$ and $4^{vw} \to 2^u$. 
In particular, the linear analysis once again only identifies causal relationships that preserve the symmetry properties of the associated structures. 
On the other hand, both the linear analysis of the POD energies (squares of POD coefficients) and the nonlinear Granger analyses additionally identify (presumably nonlinear) causal relationships from the secondary to the streamwise fluctuations that transfer between the symmetric and antisymmetric modes. In particular, this suggests that the similar behavior observed in the combined POD analysis is primarily caused by the presence and motion of the streamwise vortex in all cases. 

The causality analyses in Figure~\ref{fig:causality_split}(b)-(d) also identify several additional mechanisms that are not in the top right quadrant, such as $3^u \to 2^u$ and/or $4^u$, representing a streamwise streak translating along a wall into the corner. The reverse causal relationship, with a corner streak causing sidewall streaks (\textit{e.g.}~$4^u\to 1^u$ and $3^u$), is also observed in Figure \ref{fig:causality_split}(b) and (d).
These mechanisms are analogous to those also identified in the analysis of the combined POD modes (Figure \ref{fig:causality_full}), though here we are able to establish the extent to which such mechanisms are driven by the secondary velocity components, as opposed to the streaks themselves. In particular, the results suggest that while the secondary velocity components (streamwise vortices) are primarily responsible for the dynamics of streamwise streaks, there are nonlinear mechanisms whereby streamwise streaks affect their own dynamics.

\section{Discussion and conclusions}
\label{sec:conclusions}

In this work, we have formulated a nonlinear extension of linear Granger causality analysis that allows for the explicit consideration of nonlinear interactions (of a prescribed form) when quantifying the causal relationships between them. This methodology has been applied to study the dynamics of coherent structures present in turbulent flow through a square duct. It was demonstrated that the causal mechanisms identified are highly sensitive to both the choice of variables and the form of the model used in the analysis. 

For standard linear Granger causality applied to POD coefficients, the only causal mechanisms that are identified are between modes that share the same symmetry properties. These mechanisms govern the translation of streamwise vortices and streaks towards and away from the duct corner. Applying linear Granger causality to the POD energy contents reveals a different set of causal mechanisms, which allows for changes in the symmetry properties of the underlying modes. This is attributed to the fact that such mechanisms are agnostic to the sign of the coefficient of the cause or effect variable, so the cumulative causal effect cancels out unless only a quantity that removes dependence on the sign of the coefficient is used. The proposed nonlinear extension is able to capture both of these classes of interactions, however insight is gained when also performing linear analysis, in order to distinguish between linear and nonlinear effects. It is possible that more sophisticated methods to disambiguate between linear and nonlinear relationships could also be incorporated into this analysis workflow \cite{baddoo2022kernel}. Note that the causal effects of various linear mechanisms can also be studied by intrusively removing or inhibiting certain terms within numerical simulations \cite{lozanod2021causalitywall}.

By considering the streamwise and secondary mode components separately, we were also able to determine the causal relationships between velocity components within a single mode. It was determined that secondary velocity fluctuations are the primary causal driver of the dynamics in all cases. While the linear mechanisms associated with this are consistent with the lift-up mechanism, the nonlinear mechanisms represent a generalization of this well-known effect. 

Note that several of the analysis variants incorporated here are similar to methods used in past causality studies of turbulent flows. For example, Mart{\'\i}nez-S{\'a}nchez et al.~\cite{martinez2022causality} use POD coefficients of modes incorporating all velocity components, while Lozano-Dur{\'a}n et al.~\cite{lozano2020causality} use energy content of different streamwise and spanwise wavenumbers (thus not distinguishing the sign/direction of the underlying structures).

There are several possible directions of future work. While the models used for linear and nonlinear Granger causality analysis are here identified from data, their simple form means that they could also potentially be found from physics-based analysis, such as from projecting the linearized or full nonlinear governing equations onto identified POD modes. 
In the present work, we only presented results studying four modes at a time, which all corresponded to streamwise-invariant structures. Further analysis could also consider additional modes, including those for nonzero streamwise wavenumbers. Here the modes were computed after subtracting both the streamwise and secondary components of the mean. The causal mechanisms leading to the secondary mean flow could possibly be more directly studied by not subtracting this secondary mean component, and instead including it as a mode with time-varying coefficient.

\section*{Acknowledgments}
This work was performed in part during the Fifth Madrid Summer Workshop, funded by the European Research Council under the  Caust grant ERCAdG-101018287. HJB, STMD and BLD acknowledge support from Air Force Office of Scientific Research grant FA9550-22-1-0109.
\section*{References}

\bibliographystyle{unsrt.bst}
\bibliography{ReferencesMadrid}

\end{document}